\documentstyle[aps,graphicx,epsf]{revtex}\tighten
\draft

\def \D {\mbox{D}}

\begin{document}
\twocolumn[\hsize\textwidth\columnwidth\hsize\csname
@twocolumnfalse\endcsname

\title{Dynamics of Brane-World Cosmological Models}

\author{A. A. Coley$^{1,*}$} \address{$^1$Department of Mathematics and Statistics,
Dalhousie University, Halifax, Nova Scotia}
\maketitle

\begin{abstract}

We show that generically the  initial singularity is isotropic
in spatially homogeneous cosmological models in the brane-world scenario.
We then argue that it is plausible that the
initial singularity is isotropic
in typical brane world cosmological models.
Therefore,  brane cosmology
naturally gives rise to a set
of initial data that provide 
the conditions for inflation to subsequently take place,
thereby solving  the initial conditions problem 
and leading to a self--consistent and viable cosmology.

\end{abstract}

 \vskip2pc]

\section{Introduction}

Recent developments in string theory suggest that gravity may be
a truly higher-dimensional theory, becoming effectively
4-dimensional at lower energies. This leads to modifications to
Einstein's theory of general relativity (GR) at high 
energies and particularly at early times.
There is currently  great interest in higher-dimensional gravity theories inspired by string
theory in which the matter fields are confined to a 3-dimensional
`brane-world' embedded in $1+3+d$
dimensions, while the gravitational field can also propagate in
the $d$ extra dimensions (i.e., in the `bulk')~\cite{rubakov}. 
In this paradigm it is not
necessary for the $d$ extra dimensions to be small or even
compact, a 
departure from the standard Kaluza-Klein scenario. In recent work Randall and Sundrum~\cite{randall} have shown that for
$d=1$, gravity can be localized on a single 3-brane even when the
fifth dimension is infinite. An elegant
geometric formulation and generalization of the 
class of Randall-Sundrum-type brane-world models 
has been given in~\cite{sms}.

The dynamical
equations on the 3-brane differ from the GR
equations by terms that carry the effects of imbedding and of the
free gravitational field in the 5-dimensional bulk. 
The local (quadratic) matter
fields  corrections are
significant only at very high energies. 
In addition, there are nonlocal effects 
from the free gravitational field in the bulk, transmitted via the
projection ${\cal E}_{\mu\nu}$ of the bulk Weyl tensor, that contribute further corrections to the Einstein
equations on the brane.  ${\cal E}_{\mu\nu}$ can be irreducibly decomposed (with
respect to a timelike congruence $u^\mu$)
in terms of
an effective nonlocal energy density on the brane, ${\cal U}$, arising from the free gravitational field in the
bulk, an effective nonlocal anisotropic stress,
${\cal P}_{\mu\nu}$, on the brane, and an effective nonlocal energy flux on the brane,
${\cal Q}_\mu$
~\cite{Maartens}.

In general, the
conservation equations do not determine all of the independent components
of ${\cal E}_{\mu\nu}$ on the brane; in particular, there is no
evolution equation for ${\cal P}_{\mu\nu}$. Thus in general, the projection of the
5-dimensional field equations onto the brane does not lead to a
closed system. 
If the induced metric on the brane is flat, and the bulk is
anti-de Sitter, as in the original Randall-Sundrum scenario
\cite{randall}, then ${\cal E}_{\mu\nu}=0$. 
More importantly, in cosmology the background induced metric is not flat, but a
spatially homogeneous and isotropic Robertson-Walker (RW)
 `Friedmann' model, for which
\begin{equation}\label{6c}
\D_\mu{\cal U}={\cal Q}_{\mu}={\cal P}_{\mu\nu}=0\,,
\end{equation}
where $\D_\mu$ is the totally projected part of the brane
covariant derivative.
Hence, in such cosmological settings the evolution of ${\cal E}_{\mu\nu}$ is fully determined,
and the system of equations is closed.
In general ${\cal U}\neq0$ in the Friedmann
background. 
Much effort is currently being devoted to
understand the cosmology of the brane world scenario, and the
Friedmann brane models have been
extensively investigated~\cite{cline,BinDefLan:2000}).

Cosmological observations indicate that we live in a Universe which is
remarkably uniform on very large scales. However, the spatial homogeneity and
isotropy of the Universe is difficult to explain within the
standard GR framework since, in the presence of matter,
the class of solutions to the Einstein equations which evolve
towards a RW universe is essentially a set of measure
zero~\cite{ch73}. In the
inflationary scenario, we live in
 an isotropic region of a potentially highly 
irregular universe as the result of an expansion phase in the early universe
thereby solving many of the problems of cosmology. Thus this
scenario can successfully generate a homogeneous and
isotropic RW-like universe from initial conditions which, in the
absence of inflation, would have resulted in a universe far
removed from the one we live in today. However, still only a restricted set
of initial data will lead to smooth enough conditions for the
onset of inflation (i.e., the so-called cosmic no-hair theorems
only apply to non-generic models~\cite{js86}), so the
issue of homogenization and isotropization is still not satisfactorily solved.
Indeed, the initial conditions problem, that is to explain why  the Universe so isotropic and 
spatial homogeneous from generic initial conditions,  is one of the central problems of modern
theoretical cosmology.
We would like to revisit these issues in the context of brane cosmology.

\subsection{Governing equations}

The field equations induced on the brane, using
the Gauss-Codazzi equations, matching conditions and $Z_2$
symmetry, were derived in~\cite{sms,Maartens}, resulting in a modification of the standard GR
equations with the new terms carrying bulk effects onto the
brane:
\begin{equation}
G_{\mu\nu}=-\Lambda g_{\mu\nu}+\kappa^2
T_{\mu\nu}+\widetilde{\kappa}^4S_{\mu\nu} - {\cal E}_{\mu\nu} \equiv \kappa^2 T^{\rm tot}_{\mu\nu}\,,
\label{2}
\end{equation}
where 
$\kappa^2=8\pi/M_{\rm p}^2, \lambda \equiv 6\kappa^2/\widetilde\kappa^4$.
The generalized Friedmann equation, which determines the expansion of the universe, in the case of spatially homogeneous
cosmological models is given by
\begin{equation} 
H^2 = \frac{1}{3}\kappa^2\rho\left(1+\frac{\rho}{2\lambda}\right)
-\frac{1}{6}{}^3R+\frac{1}{3}\sigma^2+\frac{1}{3}\Lambda +
\frac{2{\cal U}}{\lambda\kappa^2}\,, \label{frie}
\end{equation}
where ${}^3R$ is the scalar curvature of the hypersurfaces orthogonal
to the fluid flow, which we associate with the cosmological
fluid,  $2\sigma^2\equiv\sigma^{ab}\sigma_{ab}$ is the
shear scalar, and
$H = {\dot a}/a $ is the mean Hubble expansion parameter.

The brane energy-momentum
tensor for a perfect fluid or a minimally-coupled scalar field is given by
\begin{equation}
T_{\mu\nu}=\rho u_\mu
u_\nu+ph_{\mu\nu}\,.
 \label{3''}
\end{equation}
where $u^\mu$ is the 4-velocity, $\rho$ and $p$ are the energy density and isotropic
pressure, and $h_{\mu\nu} \equiv g_{\mu\nu}+u_\mu u_\nu$.
It follows \cite{sms} that
the brane energy-momentum tensor separately satisfies the
conservation equation, 
\begin{equation}
\dot{\rho}+ 3H(\rho+p)=0\,.\label{cons}
\end{equation}
For a minimally coupled scalar field the energy density and pressure are, respectively,
\begin{equation}
\rho={1\over2}\dot\phi^2+ V(\phi),~ p={1\over2}\dot\phi^2-V(\phi),
\end{equation}
and Eqn (\ref{cons}) is equivalent to the Klein-Gordon equation.

There are many reasons to consider cosmological models that
are more general than RW, both spatially homogeneous and anisotropic,
and spatially inhomogeneous.

(i) The 3-curvature in RW models is given by
${}^3R = 6k / a^2$, where $k=0, \pm1$ is the curvature constant. 
An equivalent 3-curvature occurs in spatially homogeneous 
and isotropic curvature models, and a similar term occurs in other cosmological models.

(ii) For
BRW models, equation (\ref{6c}) implies that  
${\cal U}={\cal U}(t)$.
In the models of Bianchi type I, 
${\cal Q}_a=0$ but there is no restriction on 
${\cal P}_{ab}$; however, the particular case in which this term is zero can be studied.
Thus, in RW and Bianchi I models the evolution equation for ${\cal U}$ is~\cite{Maartens}
\begin{equation}
\dot{\cal
U}+ 4H{\cal U}=0\,,
\label{loccons}
\end{equation}
 which integrates to ${\cal U}={\cal U}_0 / a^4$, 
which has the structure of a `dark' radiation fluid, where 
${\cal U}_0$ can be negative.

(iii) A Bianchi~I brane is covariantly characterized in ~\cite{mss}. 
The conservation equations reduce to  Eqn (\ref{cons}), an
evolution equation for 
${\cal U}$ and a differential constraint on 
${\cal P}_{\mu\nu}$. 
The presence of ${\cal P}_{\mu\nu}$ in
the governing equations means that in general we cannot integrate
to find the shear as in GR.
However, when
the nonlocal energy density vanishes
or is negligible, i.e., ${\cal U}=0$, then the
conservation equations imply $\sigma^{\mu\nu}{\cal
P}_{\mu\nu}=0$, which is
consistent on the brane~\cite{Maartens}. 
This assumption is often made in the case of RW branes~\cite{BinDefLan:2000}.
The shear evolution equation may then be integrated to give
\begin{equation}\label{s}
\sigma^{\mu\nu}\sigma_{\mu\nu}  =
{6\Sigma^2\over a^6}\,,~~\dot\Sigma=0\,.
\end{equation}
Bianchi I models on the brane have been studied by a number of 
authors~\cite{mss,CamSop}.
A similar shear term occurs in other hypersurface orthogonal Bianchi cosmological models (such as Bianchi type V models).

(iv) We assume that the matter content is 
a non-tilting perfect fluid with a linear barotropic
equation of state, i.e., 
$p = (\gamma-1)\rho$, where the energy conditions  imply
$\rho\geq 0$, and the constant $\gamma$ satisfies $\gamma\in[0,2]$.  A dynamical analysis of scalar
field models indicates that at early times the scalar field is effectively massless.
A massless scalar field is equivalent to a perfect fluid
with a stiff equation of state parameter  $\gamma=2$. Eqn (\ref{cons}) then yields
$\rho=\rho_0 a^{-3\gamma}$, where $\rho_0 > 0$.

We can therefore write down a phenomenological generalized
Friedmann equation: 
\begin{equation}\label{f}
H^2={\kappa^2\rho_0 \over 3a^{3\gamma}}  +  {\kappa^2\rho_0^2\over 6\lambda a^{6\gamma}} - {k \over a^2} 
 + \frac{1}{3}\Lambda + {\Sigma^2\over a^6} +
{{\cal U}_0 \over a^4}.
\end{equation} 
This equation is applicable in a wide class of 
spatially homogeneous cosmological models; in particular, it
is valid in Bianchi I brane models.
In many applications the four-dimensional
cosmological constant is assumed to be zero~\cite{randall}; here we shall assume that if
it is non-zero it is positive,
i.e. $\Lambda\geq 0$. 

\section{Early Universe:}

It is of considerable  interest to study the  
classical dynamical effects in these cosmological models.
A unique feature of brane cosmology is that 
$\rho^2$ dominates at early times which will lead to completely different behaviour
to that in GR. 
First, there is the question of the existence of singularities.
The generalized Raychaudhuri equation governs
gravitational collapse and initial singularity behaviour on the
brane. The local energy density and pressure corrections
further enhance the tendency to collapse (if $2\rho+3p>0$)~\cite{mwbh}.
The nonlocal term
can act either way depending on its sign;
a negative ${\cal U}$ enhances the localization of the
gravitational field on the brane (the singularity 
can be avoided in this case), and a positive
${\cal U}$ acts against localization, and also reinforces the
tendency to collapse.
For the models governed by Eqn
(\ref{f}), it is easy to show that once a critical value of $a$ is attained, a
singularity must occur.

As in GR the powerful singularity theorems of Penrose and Hawking, which guarantee the existence 
of spacetime singularities, will be generally applicable in brane world models.
However, these theorems give little
information about the nature of the singularities they predict.
The most detailed proposal for the structure of spacetime singularities in GR
are the conjectures of Belinskii, Khalatnikov and Lifshitz (BKL)~\cite{bkl}
which essentially consist of two parts. The first is:
I. Perfect fluid spacetimes (with a {\em linear} equation of state) with  
non--stiff matter have 
the property that asymptotically
close to the singularity, matter is not dynamically significant. 
In the case of spatially homogeneous models this implies that 
spacetimes are space--like and
oscillatory  (asymptotic Mixmaster behaviour, oscillating indefinitely as
the cosmological initial singularity is approached into the past), 
or are space--like and non--oscillatory (and asymptotically Kasner at the singularity).
In the case
of stiff matter, which includes the
massless scalar field case, the matter is not insignificant near the
singularity and generically the
spacetimes have singularities which
are space--like and non--oscillatory (asymptotically of Jacobs form).

Theoretical justification for part I of the BKL conjectures in
Bianchi models, in which the Einstein equations are
a system of ODE's, has been  provided using dynamical systems methods ~\cite{WE} 
and more rigorous methods~\cite{ren2}.
There are, however, {\em special} classes of models
that do not obey the BKL conjectures in GR, the most important being models with
an {\em isotropic initial singularity\/}~\cite{GW85},
whose evolution near the cosmological initial singularity is
approximated (in an
appropriately defined  mathematical sense)
by the flat Friedmann model.

An essential feature of the brane models governed by Eqn (\ref{f}) is that 
at high densities the term
($\kappa^2\rho_0^2/ 6\lambda a^{6\gamma}$) dominates and the
the effective equation of state
becomes ultra stiff. Consequently
matter dominates the shear and curvature (and the other) terms in Eqn (\ref{f})
when $\gamma > 1$, leading to isotropic expansion of 
the early  universe in such cases. In particular, $\sigma^2/H^2 \to 0$ as
$t \to 0$.

Indeed, the flat, spatially homogeneous and isotropic non-GR brane-world
(without brane tension)
BRW model, denoted here by ${\cal F}_b$, in which $a(t) \sim t^{\frac{1}{3\gamma}}$~\cite{BinDefLan:2000}, 
is always a source/repeller for $\gamma \ge 1$ 
(as can clearly be seen from Eqn (\ref{f})).
These BRW models
are valid  as the initial singularity is approached ($t \rightarrow 0$),
and therefore
for all physically relevant values of $\gamma$
the singularity is isotropic~\cite{GW85}.
We expect this to be a generic feature of 
more general cosmological models in the brane-world scenario,
as we will discuss this further below.

From Eqn (\ref{f}),
in the brane-world scenario
anisotropy dominates at early times only for $\gamma<1$ (whereas in GR
it dominates for $\gamma<2$!), in which case  the repellers are the usual anisotropic 
Kasner models. 
In the absence of shear, the BRW models are
sources  for $\gamma \geq 1/3$ when ${\cal U} = 0$ and $\gamma \geq 2/3$ when ${\cal U} \ne 0$
\cite{CamSop}.
The fact that the initial singularity is isotropic in Bianchi type I and V models 
was noted in~\cite{mss,CamSop}.

The particular models we have considered, in which the curvature and shear are given by
the expressions in the above phenomenological Eqn (\ref{f}), are  special. In particular,
the Bianchi I and V models are not generic, and so the study of the dynamics of these models
does not shed light on the typical behaviour of spatially homogeneous brane models. In order to do
this we shall next consider the general 
Bianchi type IX (paradigm) model.

 The matter corrections to the Einstein equations on the brane are
given by
\[
S_{\mu\nu}={\textstyle{1\over12}}\rho^2
u_\mu u_\nu
+{\textstyle{1\over12}}\rho\left(\rho+2 p\right)h_{\mu\nu}\,,
\]
for a perfect fluid (or minimally-coupled scalar field). 
There are also nonlocal effects from the free gravitational
field in the bulk via ${\cal U}$.
All of the bulk corrections may be consolidated into an effective total
energy density via Eqn (\ref{2}), so that the modified Einstein equations take on the standard
perfect fluid  form with 
\[
\rho^{\rm tot} = \rho+ {\rho^2\over2\lambda} +
{\widetilde{\kappa}^{4}{\cal U}\over
\kappa^6} +
{\Lambda\over\kappa^2}, 
p^{\rm tot} = p+ {\rho^2\over2\lambda} + {\rho p\over\lambda} -
{\widetilde{\kappa}^{4}{\cal U}\over
3\kappa^6} - 
{\Lambda\over\kappa^2}\,.
\]

For the Bianchi IX models with $\gamma_{ij} = diag(a_1^2,a_2^2,a_3^2)$, where the mean scale factor
is given in terms of the three spatial scales by
$a^3 = a_1 a_2 a_3$, it can be proven~\cite{coley}
from the
conservation Eqn (\ref{cons}), 
the nonlocal conservation Eqn (\ref{loccons}),
and the perfect fluid
generalized Raychaudhuri equation that there is a singularity
(at $t=0$), and that $\rho \rightarrow \infty$ as  $t \rightarrow 0$.
Close to singularity we have that
$\dot{H} < 0$, and
\begin{equation}
\rho^{\rm tot} = \rho^{\rm eff} = {1\over2\lambda}\rho^2, ~~
p^{\rm tot} = p^{\rm eff} = {(2\gamma - 1)\over2\lambda}\rho^2.\label{bb}
\end{equation}
The Bianchi IX equations of motion then become
\begin{equation}
(\ln a_1^2)'' - (a_2^2 - a_3^2)^2 + a_1^4 = 
{\kappa^2(1 - \gamma)\over\lambda} a^6 \rho^2, et~cyc. \label{ZZZ}
\end{equation}
where a prime denotes differentiation with respect to $\tau$ (where $dt = a^3 d\tau$ so that
$\tau \rightarrow -\infty$ as the singularity is approached), and {\em et cyc.} denotes two more equations
obtained by cycling the indices of $a_i$ on the left-hand-side of Eqn (\ref{ZZZ}) (the right-hand-side
remains the same). As in the GR case, there exists a first integral
\begin{equation}
(\ln a_1^2)'(\ln a_2^2)' + 2a_1^2 a_2^2 -a_1^4 + cyc. = {\kappa^2\over8\lambda} a^6 \rho^2. \label{ZZZZ}
\end{equation}

The BRW solution ${\cal F}_b$, $a_1=a_2=a_3=a \equiv a_b$ (with 
$a_b(t) = t^{\frac{1}{3\gamma}}$), occurs as $\tau \rightarrow -\infty$ ($t \rightarrow 0$)
and $a \rightarrow \infty$.  From the governing equations
(where we use Eqn~(\ref{ZZZZ}) to eliminate $\kappa^2 \rho^2/\lambda$ in Eqns~(\ref{ZZZ})
and the Raychaudhuri equation governs the evolution of $H$), we obtain
the following 5 essential eigenvalues for the linearization about ${\cal F}_b$:
\begin{equation}
3(\gamma - 1), 3(\gamma - 1), (3\gamma -1), (3\gamma -1), (3\gamma -1)
\end{equation}
(the first two eigenvalues correspond to {\em shear} modes while the last three 
correspond to {\em curvature} modes).
Thus ${\cal F}_b$ is a past-attractor (repeller/source) of the Bianchi type IX models for $\gamma>1$ 
(for $\gamma=1$ the BRW solution is an analogue of the Jacobs stiff fluid solution in GR
and is a past-attractor). There are no other past-attractors~\cite{coley}; in particular, the analogues
of the non-flat Kasner vacuum solutions are {\em saddles}.
Hence the singularity is non-oscillatory and isotropic. It can be shown that ${\cal F}_b$ is a source
in all Bianchi models~\cite{coley}.

The results are incomplete in that
a description of the gravitational field in the
bulk is not provided. Unfortunately, the evolution of the
anisotropic stress part is {\em not} determined on the brane.
These nonlocal terms also enter into crucial dynamical equations,
such as the Raychaudhuri equation and the shear propagation
equation, and can lead to important effects. 
The correction terms must be consistently derived from the higher-dimensional equations.
Additional modifications occur for 
more general higher-dimensional (bulk) geometry, 
higher curvature corrections,
higher-dimensional matter fields in the bulk,
and for motion of the brane.

\section{Discussion}

Consequently, we have shown that {\em generically the  initial singularity is isotropic}
in spatially homogeneous brane world cosmological models.

The second part of the BKL conjectures is:
II. Each spatial point evolves towards the singularity as 
if it were a spatially homogeneous  cosmology. 
That is, generic spacetimes  have the
property that spatial points decouple near the singularity and 
the Einstein equations effectively reduce to ODEs,
so that the local
dynamical behaviour is 
asymptotically like Bianchi models near the
singularity.
Therefore, according to the BKL conjectures, the singularities in general four-dimensional 
spacetimes in GR are space--like and
oscillatory or
are space--like and non--oscillatory (e.g., for massless scalar fields).

An analysis of the behaviour of spatially inhomogeneous solutions
to Einstein's equations near an initial singularity is in its infancy,
and hence there is less support for part II of 
the BKL conjectures. However, a special class of
Abelian $G_{2}$ spatially inhomogeneous
models were analysed and it was found that
the evolution at different spatial points approach that of
different Kasner solutions ~\cite{inv}. 
A recent numerical investigation of a class of vacuum Gowdy $G_{2}$ 
cosmological spacetimes 
has shown evidence that at a generic point in space the
evolution towards the initial singularity is asymptotically that of
a spatially homogeneous spacetime with Mixmaster behavior~\cite{WIB}.
In both of these cases the presence of the
inhomogeneity ceases to govern the dynamics asymptotically toward the
singularity, thereby providing further support for the BKL conjectures.

In addition, in a recent qualitative analysis of a class of
spatially inhomogeneous $G_{2}$ brane cosmological models (with one spatial
degree of freedom) near the initial cosmological singularity, it was found that
${\cal F}_b$ is again a local source \cite{coley}. It was also argued, based upon local dynamical
considerations and physical arguments, that the main result that the singularity is isotropic will
persist when additional, more general, affects are included.

Thus it is plausible, from the  BKL conjectures in GR and from this recent
study of spatially inhomogeneous brane models, that {\em typically the
initial singularity is isotropic}
in brane world cosmological models.

Therefore, unlike the situation in GR, it is plausible that a wide range
of brane cosmological models (and of non-zero measure) admit an isotropic singularity.
Such a `quiescent' cosmology~\cite{Barrow}, 
in which the universe began in a highly 
regular state but subsequently evolved towards irregularity, might offer
an explanation of why our Universe might have began its evolution in such a smooth manner
and may provide a realisation of Penrose's ideas on gravitational entropy 
 and the second law of thermodynamics in cosmology~\cite{Penrose79}.

More importantly, it is therefore possible that a quiescent cosmological period occuring in brane
 cosmology
provides a physical scenario in which the universe starts off smooth and that naturally gives
rise to the conditions for inflation to  subsequently take place.
As noted earlier,  in the conventional scenario only a restricted 
set of initial data will lead to the conditions possible for inflation to occur.
We thus argue that brane cosmology
in tandem with  inflation 
may lead to a fully self--consistent and physically viable cosmology~\cite{GCW92}.

Subsequently, the models evolve essentially as in standard cosmology.
The modified
Friedmann equation for a flat RW brane (with $\Lambda=0={\cal
U}$) and the Klein-Gordon Eqn (\ref{cons})
yield $\dot{\rho}= -3\dot\phi^2\rho$
at early times for a scalar field source,
so that $\rho$ is monotonically decreasing and the models will eventually
evolve to the low density regime.

The intermediate dynamics is affected by the brane corrections. The issue of inflation
on the brane was investigated in~\cite{mwbh}, where it was shown
that on an RW brane in 5-dimensional anti de Sitter space
the quadratic term in $\rho$
increases friction in the inflaton field equation 
and inflation at high energies
proceeds at a higher rate than the corresponding
rate in GR.
Moreover, it was shown that, contrary to
expectations, a large initial {\em anisotropy} introduces
more damping into the scalar field equation of motion and results
in more inflation~\cite{mss}.


At late(r) times a number of features of the cosmological dynamics
can be deduced directly from Eqn (\ref{f}), consistent with the
qualitative analysis of perfect fluid RW  
and Bianchi type I and V cosmological 
models in the Randall-Sundrum brane-world scenario 
of~\cite{CamSop}.
Models with a positive curvature can recollapse. However, 
for ${\cal U} < 0$  models
can (re)collapse (even without a positive curvature) for any values of
$\gamma$. 
Indeed, for  ${\cal U}< 0$ and  positive curvature (as in the $k=1$ BRW models), 
there exist oscillating universes in which the 
physical variables oscillate periodically
without reaching any spacelike singularity~\cite{CamSop} 
(it was noted earlier that when there are bulk effects present 
a singularity can be avoided).
When a positive cosmological constant is present, the 
de Sitter model is always the global attractor for ${\cal U}\geq 0$.
For ${\cal U} < 0$,  models
can (re)collapse (even without a positive curvature)
so that in this case the  de Sitter models is only a local attractor
(and the cosmic no-hair theorem is consequently violated~\cite{VERN}).


{\em Acknowledgements:} This work was supported, in part,  by NSERC of Canada.



\end{document}